\documentclass[sigconf,nonacm]{acmart}

\usepackage{amsmath}
\usepackage{tabularx}
\usepackage{colortbl}

\AtBeginDocument{%
  }

\renewcommand\footnotetextcopyrightpermission[1]{}

\newcommand{\Dhat}{\widehat{D}}
\newcommand{\nzero}{\ensuremath{n0}}
\newcommand{\none}{\ensuremath{n1}}
\newcommand{\nthree}{\ensuremath{n3}}
\newcommand{\nseven}{\ensuremath{n7}}
\definecolor{auditnavy}{HTML}{061434}
\definecolor{auditblue}{HTML}{076DD8}
\definecolor{auditpurple}{HTML}{491080}
\definecolor{auditslate}{HTML}{52617E}
\definecolor{auditheader}{HTML}{EAF3FD}
\definecolor{auditstripe}{HTML}{F5F8FC}
\newcolumntype{Y}{>{\raggedright\arraybackslash\hyphenpenalty=10000\exhyphenpenalty=10000\relax}X}
\newcommand{\QCell}[2]{\textbf{#1}\par{\scriptsize\textcolor{auditslate}{#2}}}
\newcommand{\MTag}{\textcolor{auditblue}{\scriptsize [M/M]}}
\newcommand{\NTag}{\textcolor{auditslate}{\scriptsize [N/N]}}
\newcommand{\TrajCell}[2]{\textcolor{auditpurple}{$#1$}\par{\scriptsize\textcolor{auditslate}{#2}}}
\begin{document}

\title{Same Agent, Different Answers}
\subtitle{A Repeat-Aware Audit of Corpus-Induced Answer Churn in Retrieval-Augmented QA}

\author{Jingjie Ning}
\affiliation{%
  \institution{School of Computer Science, Carnegie Mellon University}
  \city{Pittsburgh}
  \state{Pennsylvania}
  \country{USA}
}
\email{jening@cs.cmu.edu}

\author{Xueqi Li}
\affiliation{%
  \institution{School of Computer Science, Carnegie Mellon University}
  \city{Pittsburgh}
  \state{Pennsylvania}
  \country{USA}
}
\email{xueqil@cs.cmu.edu}

\renewcommand{\shortauthors}{Ning and Li}

\begin{abstract}
A retrieval-augmented QA system can return different answers after an index expansion even when its requested model identifier, prompt, retrieval policy, evidence depth, rendering, and exposed generation controls are held fixed. Aggregate accuracy may hide these changes when gains and losses cancel, while ordinary generation variability makes one-shot comparisons overstate update effects. We call the hidden phenomenon accuracy-blind answer churn and introduce the \emph{Snapshot Compatibility Audit}, which estimates excess answer churn by subtracting same-snapshot repeat disagreement from cross-snapshot disagreement. We instantiate it by expanding one frozen FineWeb prefix from one to seven shards. In a preregistered 400-question Natural Questions study, normalized-exact and blinded-semantic excess churn are 6.44 and 10.25 percentage points while exact-match accuracy changes by only $-1.50$ points. A post-hoc analysis finds repeat-stable semantic flips on 40/400 questions. A separately preregistered 200-question TriviaQA study yields smaller, directionally consistent excess churn while exact-match accuracy moves in the opposite direction. An outcome-blind post-hoc 100-question subset replication with a second DeepSeek generator and serving configuration finds 8.75 pp of semantic excess churn even as exact match rises by 3.00 percentage points. Answer-level compatibility can therefore fail without a conspicuous or consistently directed utility shift. Retrieval-augmented releases should audit compatibility alongside utility.
\end{abstract}



\maketitle
\hypersetup{pdfauthor={Jingjie Ning and Xueqi Li}}

\begin{figure*}[t]
  \centering
  \includegraphics[width=\textwidth,keepaspectratio]{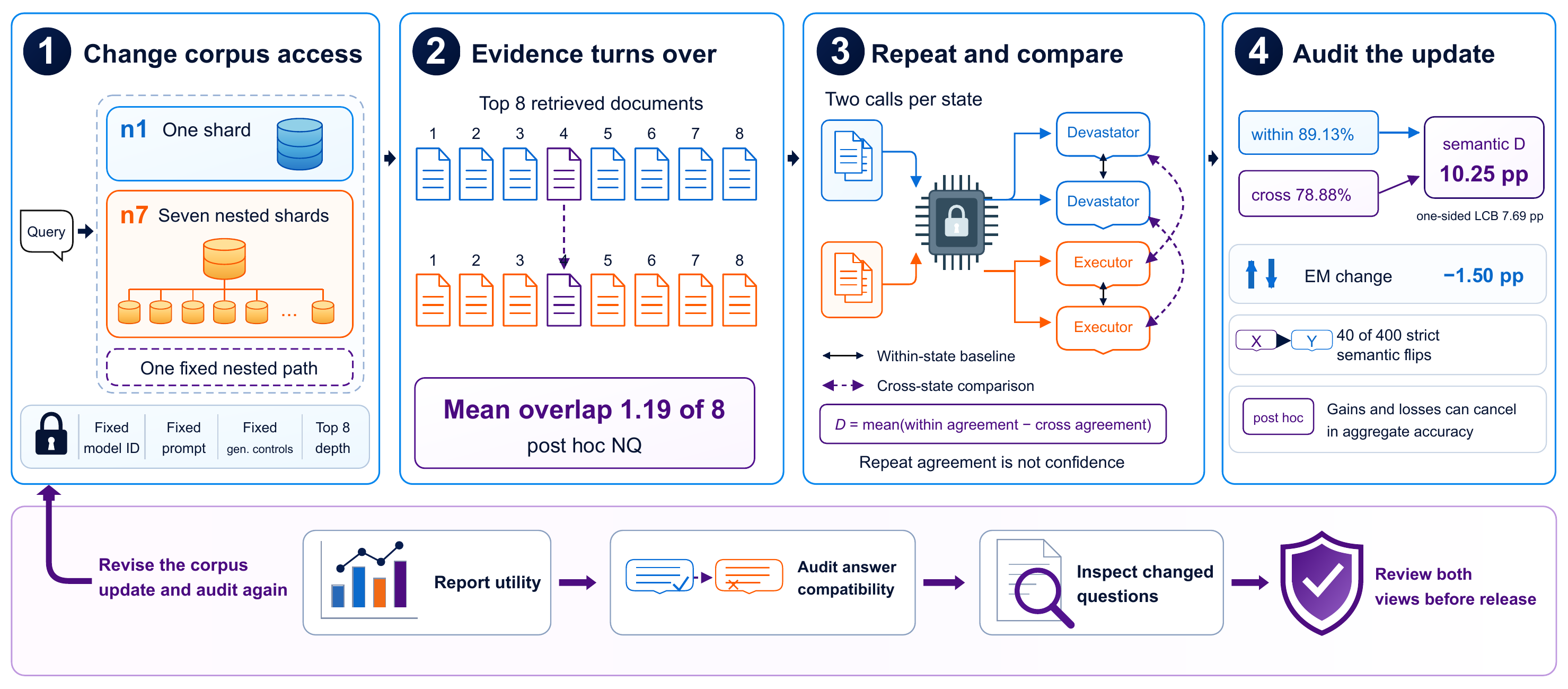}
  \caption{Overview of the Snapshot Compatibility Audit for a corpus update to a fixed single-turn retriever-generator. One nested FineWeb prefix grows from one shard to seven while all other pipeline components stay fixed. Repeated calls separate same-state variation from cross-state movement. The lower release workflow is recommended practice rather than a policy evaluated by this study. Overlap and strict semantic flips are post hoc.}
  \Description{Four numbered panels show a fixed single-turn QA system whose corpus access expands along a nested prefix, followed by top-eight evidence turnover, two repeated answers from each state, and the Snapshot Compatibility Audit. On NQ, mean endpoint top-eight overlap is 1.19 of 8, semantic excess churn is 10.25 percentage points, EM changes by minus 1.50 points, and 40 of 400 strict semantic flips are post hoc. A lower recommended workflow reports utility, audits answer compatibility, inspects changed questions, and returns to corpus revision.}
  \label{fig:teaser}
\end{figure*}

\section{Introduction}
\label{sec:introduction}

Web collections evolve, and crawling and refresh policies determine what a search system can expose \cite{cho2000evolution,cho2003effective,adar2009webchanges}. Retrieval-augmented generation couples these familiar index dynamics to generated answers. Production indexes grow, documents are refreshed, and retrieval infrastructure is rebuilt, yet such changes are usually evaluated through aggregate utility. Corpus-scaling studies likewise ask how average quality changes as more inference-time data becomes available \cite{shao2024massiveds,ning2026less}. We ask a question. \emph{Did the system preserve its answer-level behavior?}

Language-agent formulations can couple an LLM to external memory and retrieval \cite{wu2024streambench}. Here, \emph{same agent} means that the requested model identifier, prompt, retrieval policy, evidence depth, rendering, and exposed generation controls are fixed. Pairing this identity with a particular accessible corpus defines a deployed snapshot. The evaluated system is a single-turn retriever-generator rather than an autonomous multi-step policy \cite{deng2026agenticrag}.

Figure~\ref{fig:teaser} traces the full audit and shows one locked example. With one FineWeb shard, the fixed generator answered \emph{The Devastator} twice. With seven shards, it answered \emph{The Executor} twice. The Executor is plausible but refers to a different ship. Expansion can also repair an answer. For a question about who plays Jill Bigelow in \emph{Line of Duty}, two \texttt{UNKNOWN} responses become two reference-matching \emph{Polly Walker} responses. Corpus expansion therefore changes which answer regime the system selects rather than moving every question in the same utility direction.

These cases distinguish \emph{utility} from \emph{compatibility}. Utility asks whether a new snapshot is better on average. Compatibility asks whether it preserves behavior that users and downstream applications previously observed. An index rebuild can alter compatibility without changing the model binary, prompt, or public API version. This matters when answers feed caches, regression tests, automated workflows, or human decisions because balanced improvements and regressions may leave a dashboard green while changing which questions receive which facts. A useful corpus-update audit must ask whether the new state is more useful and whether it remains behaviorally interchangeable with the old one.

Aggregate accuracy is poorly suited to expose such changes. Gains and losses cancel, while transitions between two exact-match nonmatches remain invisible. Stochastic generation creates a second obstacle because a different string after an index change need not be caused by the index. The relevant baseline is how often the same system changes its answer when nothing changes. This is the RAG analogue of prediction churn and backward compatibility in model updates \cite{fard2016churn,traeuble2021backward}, although the outputs are open-ended and the model remains fixed.

We therefore compare two independent responses at each corpus scale. For a similarity function $k$, the Snapshot Compatibility Audit subtracts cross-snapshot agreement from within-snapshot repeat agreement. Positive excess answer churn $\Dhat$ means that answers move across snapshots more than they vary under repeated calls to the same snapshot. We use normalized exact agreement and a blinded semantic-equivalence judge, then perform inference by resampling whole questions. The protocol requires paired inputs and independent repeated outputs from each stochastic snapshot. Our evidence covers one concrete intervention in which a fixed nested FineWeb prefix expands from one to seven shards while the query and all other recorded interface controls remain fixed.

We study whether expanded corpus access moves answers beyond same-state generator variation in \emph{RQ1}, how much movement aggregate accuracy misses in \emph{RQ2}, and what locked post-hoc diagnostics reveal about moving questions and scale trajectories in \emph{RQ3}. Only RQ1 is confirmatory. RQ2 uses preregistered utility metrics and transitions, while RQ3 remains explicitly descriptive.

Our main study is a preregistered fresh cohort of 400 Natural Questions (NQ) \cite{kwiatkowski2019nq}. It confirms normalized-exact excess churn of 6.44 percentage points (pp) and semantic excess churn of 10.25 pp, while aggregate exact match (EM) moves by only $-1.50$ pp. A separately preregistered 200-question TriviaQA study \cite{joshi2017triviaqa} gives smaller but directionally consistent excess churn even though its EM moves in the opposite direction, $+1.25$ pp. A post-hoc decomposition finds strict repeat-stable semantic flips on 40/400 NQ and 5/200 TriviaQA questions. An outcome-blind 100-question NQ subset replication with a second DeepSeek generator and serving configuration yields 8.75 pp semantic excess churn while EM rises by 3.00 pp.

We make three contributions.
\begin{itemize}
\item We identify \emph{accuracy-blind answer churn}. Expanding the accessible corpus can make a new retrieval-augmented snapshot behaviorally incompatible with the earlier snapshot even when aggregate utility barely moves.
\item We introduce the \emph{Snapshot Compatibility Audit}, a repeat-aware protocol for comparing stochastic system snapshots. Its excess-answer-churn estimand accounts for same-snapshot repeat variability. Whole-question inference and flip diagnostics reveal behavior hidden by aggregate utility.
\item We provide preregistered evidence on NQ, supportive cross-benchmark evidence on TriviaQA, and locked post-hoc analyses spanning a second within-family generator configuration, evidence turnover, scale trajectories, and cross-family semantic-judge robustness.
\end{itemize}

\section{Related Work}
\label{sec:related}

\subsection{Retrieval Architectures and Corpus Scaling}

Retrieval-based language models externalize knowledge through sparse or dense search, passage-fusion readers, and non-parametric stores \cite{guu2020realm,karpukhin2020dpr,lewis2020rag,petroni2021kilt, khandelwal2020knnlm,izacard2021fid}. Later systems retrieve from trillion-token stores, expose replaceable indexes, or wrap frozen black-box generators \cite{borgeaud2022retro,izacard2023atlas,ram2023incontext,shi2024replug}. Active and selective systems decide when or how much evidence to use \cite{jiang2023flare,xu2024recomp,asai2024selfrag}, while Chain-of-Note generates per-document reading notes to assess noisy or irrelevant evidence \cite{yu2024chainnote} and empirical analysis shows retrieval can help long-tail facts yet mislead elsewhere \cite{mallen2023trust}. Retrieval-pretraining, datastore/corpus-scaling, and agentic rollout-breadth studies primarily report aggregate utility \cite{wang2023retrievalpretrain,shao2024massiveds,ning2026less,xu2026ragwild,murali2026beyond}; rollout breadth can also retrieve overlapping evidence from redundant initial queries. We fix the query, generator, and retrieval depth, change corpus access, and test compatibility along one nested path.

\subsection{Evidence Robustness and RAG Evaluation}

Retrieved or supplied context can degrade QA when it is irrelevant, detrimental, position-sensitive, adversarial, or conflicting \cite{shi2023distracted,oh2023detrimental,liu2024lost,yoran2024robust, hong2024gullible,park2024robust,shen2024implicit}. Conflict studies examine whether answers follow supplied context or previously elicited knowledge \cite{longpre2021conflicts,xie2024chameleon}, while DisentQA explicitly separates parametric and contextual predictions \cite{neeman2023disentqa}. Context-faithful prompting offers a complementary mitigation objective \cite{zhou2023contextfaithful}. Our no-retrieval output is only a descriptive closed-book anchor, not a direct observation of parametric memory. RGB tests noise robustness, negative rejection, information integration, and counterfactual robustness \cite{chen2024rgb}. RAGAS, ARES, RAGChecker, and RAGTruth collectively evaluate relevance, faithfulness, component errors, and hallucinations \cite{es2024ragas,saadfalcon2024ares,ru2024ragchecker,niu2024ragtruth}. Prior work measures answer consistency under constructed context changes and robustness to an added retrieved document \cite{longpre2021conflicts,shaier2024desiderata,park2024robust}. Factual-consistency work varies semantically equivalent question forms and retrieval augmentation, Con-RAG decomposes retriever, generator, and end-to-end consistency across paraphrased queries, and Stable-RAG measures answer sensitivity to permutations of a fixed retrieved set \cite{hagstrom2023factualconsistency,hamman2025conrag,zhang2026stablerag}. We fix the query, expand a nested corpus, let evidence change endogenously, and estimate snapshot-to-snapshot semantic excess churn after subtracting same-snapshot disagreement.

\subsection{Behavioral Compatibility and Stochastic Evaluation}

Prediction churn and backward-compatible updating show that aggregate performance can conceal instance-level changes and negative flips \cite{fard2016churn,bansal2019updates,bahri2021churn,yan2021positive, traeuble2021backward}.  The concern extends to structured NLP, data updates, deployed language systems, and generative LLM evolution \cite{cai2022regression,caciolai2023regression,schumann2024backward, echterhoff2024muscle}; unlike those works, we change the retrieval corpus rather than model weights.  Hosted LLM behavior can drift across service snapshots and nominally repeated calls \cite{chen2023chatgptdrift,atil2025nondeterminism}.  Self-consistency is also fragile when questions admit multiple valid readings \cite{bartsch2023selfconsistency}, motivating uncertainty-aware evaluation \cite{miller2024errorbars,madaan2024variance,lior2025reliableeval}. Our within-versus-cross contrast is related to kernel two-sample testing \cite{gretton2012mmd}.  LLM judges enable scalable open-ended evaluation \cite{zheng2023llmjudge,liu2023geval}, but position and self-preference biases remain \cite{zheng2023llmjudge,wang2024fairevaluators, panickssery2024selfpreference}; this motivates our blinding and cross-family audit.

\subsection{Dynamic Corpora and Agent Knowledge Interfaces}

Web evolution, crawling, and refresh work treats the indexed collection as a time-varying system object \cite{cho2000evolution,cho2003effective,fetterly2003large, ntoulas2004whatsnew,adar2009webchanges}. SituatedQA and time-aware language modeling formalize extra-linguistic context and changing facts \cite{zhang2021situatedqa,dhingra2022timeaware}. StreamingQA, RealTime QA, and FreshLLMs target news streams, recurring current-world questions, or fresh QA \cite{liska2022streamingqa,kasai2023realtimeqa,vu2024freshllms}. Our frozen experiment uses FineWeb \cite{penedo2024fineweb} but is not a temporal refresh. These works motivate snapshot-aware deployment tests but do not instantiate our treatment. Prior work distinguishes fixed RAG from systems that iteratively refine evidence, while interaction logs show later agent queries often reuse terms from accumulated evidence \cite{deng2026agenticrag,ning2026agenticsearch}. A difference at one retrieval interface may propagate through a trajectory or be corrected downstream. Measuring trajectory-level compatibility remains open, and our single-turn estimand is neither an estimate nor a lower bound for it. Ambiguous questions can support multiple interpretations \cite{min2020ambigqa}. Language-agent architectures store and retrieve experience or feedback \cite{park2023generative,shinn2023reflexion,zhang2023semiparametric}. HippoRAG and LongMemEval broaden this line to long-term knowledge retrieval and conversational-memory evaluation \cite{gutierrez2024hipporag,wu2025longmemeval}. Applying our protocol there is a proposed extension rather than an empirical claim.

\section{Measuring Corpus-Induced Churn}
\label{sec:method}

\subsection{Snapshot Excess-Churn Estimand}
\label{sec:estimand}

\paragraph{Scope.} Our testbed is a fixed single-turn retriever-generator. It does not plan, choose tools, reformulate queries, retrieve iteratively, write memory, or interact across turns. The empirical claims concern corpus access along one frozen shard path. The audit itself is interface-level and can compare stochastic snapshots, but it attributes only the changes isolated by a given comparison.

Let $L$ and $H$ denote two deployed snapshots evaluated on the same questions. In our experiment, they share the requested model identifier, prompt, retriever, evidence depth, rendering, and exposed generation controls and differ only in accessible corpus scale. For question $i$, we draw two independent generator responses at each state, $L_{i,a},L_{i,b}$ and $H_{i,a},H_{i,b}$. Question-state evidence is locked, so repeats measure generator rather than retrieval variability. We define the following quantities for similarity $k\in[0,1]$.
\begin{align}
 w_i &= \tfrac{1}{2}\left[k(L_{i,a},L_{i,b}) +
                         k(H_{i,a},H_{i,b})\right],\\
 c_i &= \tfrac{1}{4}\sum_{r\in\{a,b\}}\sum_{s\in\{a,b\}}
                         k(L_{i,r},H_{i,s}),\\
 \Dhat &= \tfrac{1}{N}\sum_{i=1}^{N}(w_i-c_i).
 \label{eq:d}
\end{align}
Thus $\Dhat>0$ means that cross-snapshot answers are less similar than ordinary same-snapshot repeats. We call $\Dhat$ \emph{excess answer churn}, the increase in cross-snapshot disagreement above the average same-snapshot disagreement floor. It is an agreement gap rather than the percentage of questions whose answer changed. In this experiment, it is a path-specific corpus-associated contrast rather than a universal scaling effect.

The noise correction can be written equivalently as follows.
\begin{equation}
 \Dhat=(1-\bar c)-(1-\bar w).
 \label{eq:excess-disagreement}
\end{equation}
The first term is observed cross-state disagreement and the second is the same-state noise floor.  A one-answer-per-state comparison estimates only the first and consequently attributes ordinary decoding variation to the update.

We use two co-primary kernels.  The \emph{normalized-exact} kernel is one when two answers match after the benchmark normalization and zero otherwise.  The \emph{semantic} kernel is a blinded pairwise judgment of whether two strings give the same factual answer under reasonable aliases.  Each question's eight answers across four scales are anonymously labeled, and all 28 unordered pairs are judged before gold answers are loaded.

\subsection{What the Contrast Measures}

Let $P_i^L$ and $P_i^H$ denote the conditional answer distributions induced by the two snapshots.  With independent draws within each distribution, the population version of the question-level contrast is
\begin{equation}
 \begin{aligned}
 D_i(k)=\tfrac{1}{2}\big[&\mathbb{E}_{X,X'\sim P_i^L}k(X,X')+
 \mathbb{E}_{Y,Y'\sim P_i^H}k(Y,Y')\big]\\
 &-\mathbb{E}_{X\sim P_i^L,Y\sim P_i^H}k(X,Y),
 \end{aligned}
 \label{eq:population-d}
\end{equation}
The observed $w_i-c_i$ is an unbiased two-repeat estimate of Equation~\ref{eq:population-d}. Two repeats are the minimum that let us estimate same-state agreement without self-comparisons, an important distinction when nominally identical LLM calls need not reproduce exactly \cite{atil2025nondeterminism}. Because $k\in[0,1]$, the sample contrast lies in $[-1,1]$. Negative finite-sample values are retained rather than clipped.

For normalized equality, let $p_{i,L}(y)$ and $p_{i,H}(y)$ be the probabilities of normalized answer $y$.  Equation~\ref{eq:population-d} becomes
\begin{equation}
 D_i(k_{\mathrm{exact}})=\tfrac{1}{2}\sum_y
       \left(p_{i,L}(y)-p_{i,H}(y)\right)^2.
 \label{eq:exact-population}
\end{equation}
It is zero when the two conditional answer distributions coincide and positive when probability mass is redistributed, even when their expected EM scores are equal.

For a positive-semidefinite kernel, $2D_i(k)$ is the squared maximum mean discrepancy between $P_i^L$ and $P_i^H$, and twice Equation~\ref{eq:d} is its unbiased two-sample estimator \cite{gretton2012mmd}. The exact-equality kernel satisfies this condition. We do not assume that pairwise LLM semantic judgments form a positive-semidefinite kernel. Rare non-transitive judgments are retained and audited, so semantic $\Dhat$ is interpreted operationally as an agreement gap rather than an unconditional MMD estimate.

The subtraction is consequential.  On NQ, cross-scale semantic disagreement is 21.125\%, but ordinary same-scale repeat disagreement is already 10.875\% so their difference is the 10.250-pp semantic $\Dhat$.  Under normalized exact matching, the raw cross-scale disagreement is 81.188\%, but 74.750\% remains when the corpus state is unchanged, leaving 6.438 pp.  A raw ``answer changed'' indicator would therefore substantially overstate corpus-induced movement. Conversely, a strict stable flip contributes the maximum question-level value of one, while partial or offsetting relation changes contribute fractional or negative values.

\subsection{Inference and Confirmatory Gate}

Inference resamples whole questions, retaining all eight outputs and 28 semantic relations for a sampled question.  Each study uses a preregistered 50,000-draw bootstrap stream.  The NQ decision gate requires normalized-exact $\Dhat\geq3$ pp, a positive one-sided 95\% lower confidence bound (LCB) for exact $\Dhat$, and a positive one-sided 95\% LCB for semantic $\Dhat$. This is an intersection-union rule under which no secondary scale, subgroup, F1 score, or accuracy result can rescue a failed gate. We also report two-sided intervals. TriviaQA is a frozen supportive validation, never pooled with NQ.

\subsection{Post-Hoc Stable-Flip Diagnostic}
\label{sec:flipdef}

After the primary analyses were locked, we defined a stricter question-level diagnostic. A question is a \emph{repeat-stable semantic flip} between $L$ and $H$ when
\begin{equation}
 k(L_a,L_b)=k(H_a,H_b)=1,\qquad
 k(L_r,H_s)=0\ \ \forall r,s\in\{a,b\}.
 \label{eq:flip}
\end{equation}
``Stable'' here means semantic agreement across two calls, not verbatim identity, calibrated confidence, or factual correctness. This diagnostic is post-hoc and explains the locked estimand. Equation~\ref{eq:d} remains the inferential backbone.

\section{Experimental Design}
\label{sec:design}

\subsection{Data and Corpus Treatment}

The confirmatory population is a fresh hash-selected set of 400 questions from a 3,610-question NQ test artifact.  We excluded all 2,400 question IDs used in prior development, then took the first 400 under a preregistered SHA-256 order (seed \texttt{20260810}), without answerability, topic, retrieval, or outcome filtering.  The supportive population is an independent hash-selected set of 200 questions from the 9,951-question TriviaQA RC-Web validation split, using seed \texttt{20260813}.  Questions and selection hashes, but not gold aliases, were frozen before retrieval.

The search service uses the reproducible DeepResearchGym infrastructure \cite{coelho2026deepresearchgym} over a FineWeb index \cite{penedo2024fineweb}. Before the experiment, that index had been partitioned into seven shards. Its search interface exposes only fixed nested prefixes. We evaluate no retrieval ($\nzero$) and prefixes of one, three, and seven shards ($\none,\nthree,\nseven$). This is an accessible-scale treatment rather than a shard-order experiment. The relation $\none\subset\nthree\subset\nseven$ defines a single frozen path. At each nonzero scale, the unchanged question retrieves the top eight documents. Each rendered document includes its identifier, URL, and up to 1,200 characters of whitespace-compacted text. Rendering and prompt surface are fixed.

The answer prompt asks for a concise answer using the model's own knowledge and any supplied evidence, warns that evidence may be incomplete, irrelevant, or untrusted, and permits the exact response \texttt{UNKNOWN}.  Each independent call must return one nonempty structured answer of at most 512 UTF-8 bytes. These constraints are identical across scales and benchmarks.

\begin{table*}[t]
  \caption{Locked experimental design. NQ is confirmatory, while TriviaQA is a separately preregistered supportive validation.}
  \label{tab:design}
  \small
  \begin{tabularx}{\textwidth}{lXX}
    \toprule
    \rowcolor{auditheader}
    \textcolor{auditnavy}{\textbf{Component}} & \textcolor{auditnavy}{\textbf{NQ}} & \textcolor{auditnavy}{\textbf{TriviaQA}} \\
    \midrule
    Cohort & 400 fresh questions after excluding 2,400 development IDs &
    200 RC-Web validation questions \\
    \rowcolor{auditstripe}
    Conditions & \multicolumn{2}{c}{$\nzero,\none,\nthree,\nseven$; top-8 evidence at nonzero scales} \\
    Retrieval audit & 1,200 primary + 1,200 audit cells; 100\% exact &
    600 primary + 600 audit cells; 100\% exact \\
    \rowcolor{auditstripe}
    Generator & \multicolumn{2}{c}{\texttt{deepseek-v4-flash}; tools off; independent singleton sessions} \\
    Answer cells & 400 $\times$ 4 $\times$ 2 = 3,200 &
    200 $\times$ 4 $\times$ 2 = 1,600 \\
    \rowcolor{auditstripe}
    Semantic judging & 400 blind calls, 28 pairs/question &
    200 blind calls, 28 pairs/question \\
    Inference & 50,000 whole-question bootstrap draws &
    50,000 whole-question bootstrap draws \\
    \bottomrule
  \end{tabularx}
\end{table*}

\subsection{Locked Generation and Delayed Gold}

We collected primary and audit retrievals into independent empty caches before generation. All NQ and TriviaQA retrieval cells, including every endpoint, matched exactly in ordered document IDs and rendered evidence bytes across the two passes. Generation used \texttt{deepseek-v4-flash} through a Claude CLI harness configured to route calls through the DeepSeek adapter in bare singleton mode, requesting low reasoning effort, with tools disabled, no session persistence, and a fixed JSON output schema. Temperature and top-p were not set by our harness, and their numerical provider-default values are not recorded. For each question and scale, the same prompt and locked evidence were sent twice, producing 4,800 accepted answers. We used Williams orders \cite{williams1949design} to counterbalance the four conditions, and the second repeat reversed the first repeat's order. \texttt{UNKNOWN} and EM-nonmatching outputs were accepted without selective redraw.

The protocol used four successive gates covering the retrieval lock, answer lock, blind semantic lock, and gold unlock. Durable attempt records were written before external calls; successful cells were immutable; and only missing or transport-invalid cells could resume.  The semantic judge saw the question and eight anonymous answers, but not scale, repeat, evidence, document IDs, gold, or correctness. Gold aliases were parsed only after both answer and semantic ledgers validated. The protocol, selection, schedules, prompts, source files, model controls, analysis code, and bootstrap streams were hash-bound before the relevant call stage.

After locking the primary analyses, we selected an outcome-blind hash sample of 100 questions from the frozen NQ-400 cohort and reran the frozen $\none$ and $\nseven$ prompts and evidence twice through a stateless direct \texttt{deepseek-v4-pro} interface. Thinking was enabled, the request submitted low reasoning effort, tools were absent, and temperature and top-p were not set. All 400 answers and 100 blind V4-Flash judgments covering all six endpoint pairs per question were locked before gold access. Because the requested model and serving interface both changed and provider-default sampling values were not recorded, this is a post-hoc within-family generator-and-serving-configuration replication rather than a model-only or cross-family comparison.

\subsection{Metrics and Robustness Checks}

We report normalized EM, token F1, and \texttt{UNKNOWN} rate at all scales. NQ and TriviaQA use their respective official-style normalizations and all available aliases.  Correctness transitions are labeled \emph{EM-match} and \emph{EM-nonmatch}, rather than correct and wrong, because long answers that contain a reference can fail full-string EM.

Post-hoc diagnostics use only frozen outputs and labels. We compare exact document-ID overlap between the $\none$ and $\nseven$ top-eight sets, decompose stable flips across the scale ladder, and compare endpoint answers with the repeat-stable $\nzero$ response. Finally, an outcome-blind hash sample of 50 questions (34 NQ, 16 TriviaQA; 1,400 pair labels) was independently judged by OpenAI \texttt{gpt-5.6-sol} after the primary analysis was locked. This checks semantic-judge family robustness and is not human validation. The generator replication above and this cross-family judge audit address different sources of sensitivity.

\subsection{Reproducibility Artifact}

The anonymized artifact contains ID-only manifests, sanitized answer and judgment cells and attempts, terminal locks and results, code, summaries, and a claim matrix. It excludes question text, gold aliases, retrieved passages, and provider traces while retaining their hashes. It covers 5,200 answers, 17,400 DeepSeek labels, and 1,400 OpenAI labels. NQ, TriviaQA, and V4-Pro have 3,272/1,641/407 answer attempts and 403/204/101 judge attempts. A zero-network verifier checks completeness and regenerates churn, transitions, stability, overlap, cross-judge, V4-Pro, and trajectory claims. Gold-dependent EM and F1 stay locked.

\section{Results}
\label{sec:results}

\subsection{Expanded Corpus Access Moves Answers Beyond Repeat Noise}

The preregistered endpoint test confirms that expanding access from $\none$ to $\nseven$ moves NQ answers beyond ordinary repeat variation. Normalized-exact agreement falls from 25.25\% within snapshots to 18.81\% across snapshots, leaving $\Dhat=6.44$ pp of excess answer churn. Its one-sided 95\% LCB is 4.56 pp, and the point estimate exceeds the preregistered 3-pp threshold. Semantic agreement falls from 89.13\% to 78.88\%, yielding $\Dhat=10.25$ pp with a 7.69-pp LCB. NQ therefore passes all three confirmatory gates in Table~\ref{tab:primary}. The semantic effect persists after alias and formatting normalization, making a purely surface-form explanation unlikely.

\begin{table*}[t]
  \caption{Preregistered $\none\rightarrow\nseven$ excess answer churn. All values are percentage points. LCB is the one-sided 95\% whole-question bootstrap lower bound. It enters the frozen NQ decision rule and is reported descriptively for TriviaQA, while CI is the two-sided 95\% interval.}
  \label{tab:primary}
  \small
  \begin{tabular}{llrrrrr}
    \toprule
    \rowcolor{auditheader}
    \textcolor{auditnavy}{\textbf{Study}} & \textcolor{auditnavy}{\textbf{Kernel}} & \textcolor{auditnavy}{\textbf{Within}} & \textcolor{auditnavy}{\textbf{Cross}} & \textcolor{auditnavy}{\textbf{$\Dhat$}} & \textcolor{auditnavy}{\textbf{LCB}} & \textcolor{auditnavy}{\textbf{95\% CI}} \\
    \midrule
    NQ confirmatory & Normalized exact & 25.250 & 18.813 & \textbf{6.438} & \textbf{4.563} & [4.188, 8.750] \\
    \rowcolor{auditstripe}
    NQ confirmatory & Blind semantic & 89.125 & 78.875 & \textbf{10.250} & \textbf{7.688} & [7.188, 13.438] \\
    TriviaQA supportive & Normalized exact & 64.500 & 61.500 & \textbf{3.000} & \textbf{0.500} & [0.000, 6.250] \\
    \rowcolor{auditstripe}
    TriviaQA supportive & Blind semantic & 96.750 & 94.625 & \textbf{2.125} & \textbf{0.375} & [0.125, 4.500] \\
    \bottomrule
  \end{tabular}
\end{table*}

TriviaQA provides a directionally consistent supportive result at a different accuracy level and smaller magnitude. Its normalized-exact and semantic excess-answer-churn estimates are 3.00 and 2.125 pp, with one-sided LCBs of 0.50 and 0.375 pp. The two-sided normalized-exact interval touches zero. TriviaQA remains a separately preregistered supportive validation and is not pooled with NQ. Frozen bootstrap draws with $\Dhat\leq0$ account for 0\% under both NQ kernels, 2.764\% under TriviaQA exact, and 2.174\% under TriviaQA semantic.

A locked post-hoc decomposition shows no high-scale collapse in pairwise generator stability. At the NQ endpoints, exact within-snapshot agreement is 25.25\% at both scales and semantic agreement is 88.50\% versus 89.75\%; TriviaQA semantic agreement is 97.00\% versus 96.50\%.

\subsection{A Second Configuration Reproduces Churn}

The outcome-blind post-hoc NQ-100 subset also finds movement beyond repeat noise. V4-Pro normalized-exact and semantic $\Dhat$ are 7.25 and 8.75 pp, with one-sided LCBs of 3.25 and 4.75 pp and two-sided intervals of [2.50, 12.50] and [4.00, 14.25] pp. Six questions are strict semantic flips. EM rises from 38.00\% to 41.00\%, with a $+3.00$-pp interval of [$-2.00$, 8.50] pp. On the same question IDs, frozen V4-Flash exact and semantic $\Dhat$ are 6.25 and 9.25 pp with zero EM change. Paired V4-Pro-minus-V4-Flash intervals are [$-4.75$, 6.75] pp for exact and [$-6.75$, 5.75] pp for semantic, both including zero. The matched semantic comparison remains descriptive because frozen V4-Flash labels came from eight-answer, 28-pair packets, whereas V4-Pro labels came from four-answer, six-pair endpoint packets. Churn directionally reproduces while the endpoint EM point estimate moves opposite to the main NQ result. This replication is reported separately.

\begin{figure}[t]
  \centering
  \includegraphics[width=\columnwidth]{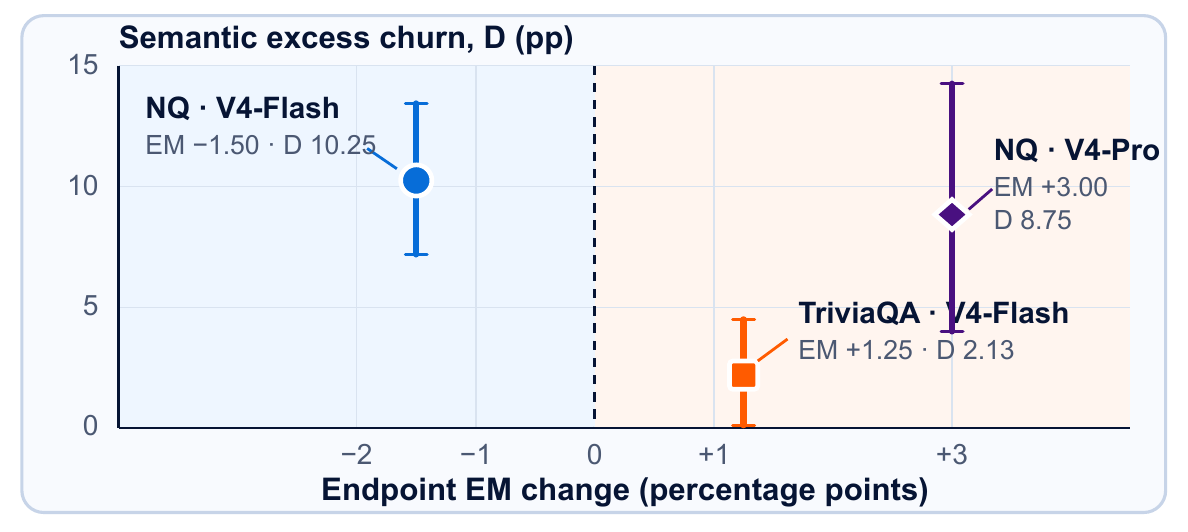}
  \caption{Semantic excess churn stays positive as endpoint exact match falls or rises. Points show locked V4-Flash NQ and TriviaQA estimates and post-hoc V4-Pro NQ; whiskers show two-sided 95\% intervals.}
  \Description{A scatter plot places endpoint exact-match change on the horizontal axis and semantic excess answer churn on the vertical axis. Natural Questions with V4-Flash has negative exact-match change and positive churn, while TriviaQA with V4-Flash and Natural Questions with V4-Pro have positive exact-match changes and positive churn.}
  \label{fig:churn-utility}
\end{figure}

\subsection{Accuracy Is an Incomplete Dashboard}

Figure~\ref{fig:churn-utility} exposes the distinction between utility and compatibility. From $\none$ to $\nseven$, NQ EM decreases by 1.50 pp, TriviaQA EM increases by 1.25 pp, and V4-Pro EM increases by 3.00 pp, while semantic excess churn remains positive in all three settings. Closed-book EM is highest at every observed scale, including 17.625\% versus 15.125\% at $\nseven$ for NQ and 68.250\% versus 63.750\% for TriviaQA; generic FineWeb retrieval therefore shows no utility gain here.

\begin{figure*}[t]
  \centering
  \includegraphics[width=0.83\textwidth]{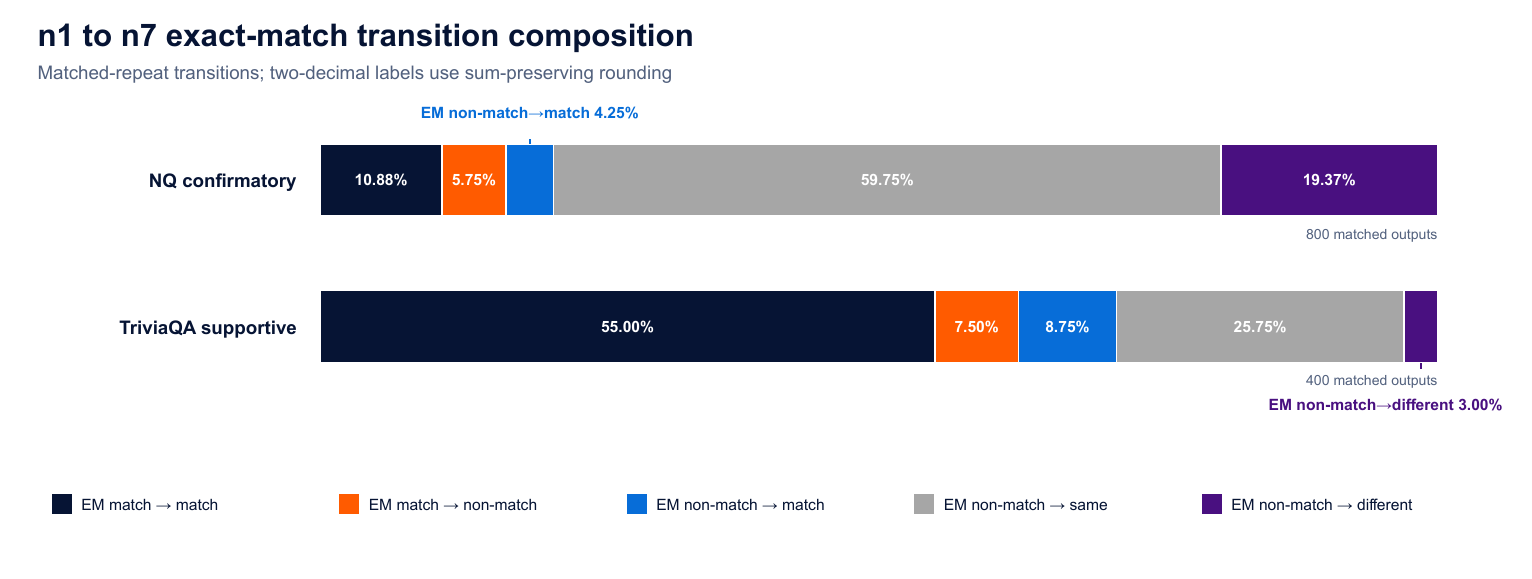}
  \caption{Matched-repeat EM transition composition from $\none$ to $\nseven$.  Aggregate EM retains only the difference between nonmatch$\to$match and match$\to$nonmatch flows; it discards their gross volume and every different-answer transition inside the EM-nonmatch class.}
  \Description{Two stacked bars decompose 800 Natural Questions comparisons and 400 TriviaQA comparisons. Natural Questions has 5.8 percent match-to-nonmatch, 4.25 percent nonmatch-to-match, and 19.4 percent nonmatch-to-a-different-nonmatch. TriviaQA has 7.5, 8.8, and 3.0 percent in the corresponding categories.}
  \label{fig:transitions}
\end{figure*}

For binary EM, the cancellation is
\begin{equation}
 \Delta\mathrm{EM}=\Pr(\text{nonmatch}\!\to\!\text{match})-
                    \Pr(\text{match}\!\to\!\text{nonmatch}).
 \label{eq:accuracy-cancellation}
\end{equation}
Accuracy subtracts the two directional flows. Compatibility cares that both occur and additionally includes changes between two EM-nonmatching answers. The matched-repeat decomposition appears in Figure~\ref{fig:transitions}. On NQ, 46/800 comparisons move from EM-match to EM-nonmatch and 34/800 move in the reverse direction.  Their difference is only $-12/800=-1.50$ pp, but the gross correctness flow is 80/800 (10.00\%).  A further 155/800 (19.38\%) move between semantically different EM-nonmatches and cannot affect binary accuracy at all. TriviaQA similarly has 30 losses and 35 gains, whose difference produces $+1.25$ pp, plus 12/400 (3.00\%) different-nonmatch transitions.  Most changes are between non-\texttt{UNKNOWN} answers, so the pattern is not simply greater willingness to answer.

These transitions are diagnostic rather than noise-corrected. A matched pair can differ partly because generation is stochastic. They show how a net score discards behavior, while $\Dhat$ determines whether cross-state movement exceeds the same-state baseline.

\subsection{Repeat-Stable Flips Explain Churn}
\label{sec:stable}

Forty of 400 NQ questions and 5/200 TriviaQA questions meet the strict criterion in Equation~\ref{eq:flip}. Each contributes the maximum per-question value of one to semantic $w_i-c_i$. Consequently, the 40 NQ flips contribute 10.00 pp of the total 10.25 pp semantic $\Dhat$, while all other questions net only $+0.25$ pp. The five TriviaQA flips contribute 2.50 pp, while the remainder nets $-0.375$ pp. This diagnostic explains the locked effect without replacing its preregistered endpoint test.

The diagnostic is semantic rather than literal. Only 2/40 NQ flips have verbatim-identical repeats at both scales, and none of the five TriviaQA flips do. All four endpoint outputs are EM-nonmatches for 35/40 NQ flips, so most of these flips are completely invisible to binary EM. The raw-string flip set need not be semantically different. We therefore avoid ``confidence'' and use \emph{repeat-stable} only in the two-call semantic sense.

Table~\ref{tab:flip-examples} shows why a single utility delta is insufficient. Corpus expansion can remove an EM-matching answer, recover one from abstention, select a different interpretation of an underspecified question, or traverse a non-monotone $X\to Y\to X$ trajectory while EM remains unchanged throughout.

\begin{table*}[t]
  \caption{Illustrative post-hoc NQ answer trajectories from the frozen ledger.}
  \label{tab:flip-examples}
  \small
  \begingroup
  \renewcommand{\arraystretch}{1.06}
  \setlength{\tabcolsep}{4pt}
  \begin{tabularx}{\textwidth}{@{}>{\raggedright\arraybackslash\hyphenpenalty=10000\exhyphenpenalty=10000\relax}p{0.17\textwidth}YYY>{\centering\arraybackslash}p{0.12\textwidth}@{}}
    \toprule
    \rowcolor{auditheader}
    \textcolor{auditnavy}{\textbf{Question and behavior}} & \textcolor{auditnavy}{\textbf{$\none$ answers a/b}} & \textcolor{auditnavy}{\textbf{$\nthree$ answers a/b}} &
    \textcolor{auditnavy}{\textbf{$\nseven$ answers a/b}} & \textcolor{auditnavy}{\textbf{Semantic path}} \\
    \midrule
    \QCell{Darth Vader's Star Destroyer}{EM regression} &
    \emph{Devastator} / \emph{Devastator} \MTag &
    \emph{Executor} / \emph{Executor} \NTag &
    \emph{Executor} / \emph{Executor} \NTag &
    \TrajCell{X\to Y\to Y}{endpoint flip} \\
    \rowcolor{auditstripe}
    \QCell{Who plays Jill Bigelow?}{abstention recovery} &
    \texttt{UNKNOWN} / \texttt{UNKNOWN} \NTag &
    \texttt{UNKNOWN} / \texttt{UNKNOWN} \NTag &
    \emph{Polly Walker} / \emph{Polly Walker} \MTag &
    \TrajCell{X\to X\to Y}{endpoint flip} \\
    \QCell{Grand National top three}{underspecified year} &
    2013 winners, Auroras Encore, Cappa Bleu, Teaforthree \NTag &
    2019 winners, Tiger Roll, Magic of Light, Rathvinden \NTag &
    2019 winners, Tiger Roll, Magic of Light, Rathvinden \NTag &
    \TrajCell{X\to Y\to Y}{endpoint flip} \\
    \rowcolor{auditstripe}
    \QCell{When did Rachel have her baby?}{EM blind spot} &
    April 4, 2002 / April 4, 2002 \NTag &
    May 16, 2002 / May 16, 2002 \NTag &
    May 16, 2002 / May 16, 2002 \NTag &
    \TrajCell{X\to Y\to Y}{endpoint flip} \\
    \QCell{New \emph{Gotham} season release?}{non-monotone} &
    final season in January 2019 / same \NTag &
    no new season; series ended April 2019 / same \NTag &
    final season premiered January 3, 2019 / same \NTag &
    \TrajCell{X\to Y\to X}{adjacent flips} \\
    \bottomrule
  \end{tabularx}
  \vspace{2pt}
  {\footnotesize\textit{Reading key.} Each scale cell gives faithful answer-bearing spans for repeats a/b. Bracketed M/N markers record normalized EM match status for the complete outputs. $X,Y$ denote locked gold-blind semantic classes, not confidence.}
  \endgroup
\end{table*}

\begin{figure*}[t]
  \centering
  \includegraphics[width=0.92\textwidth]{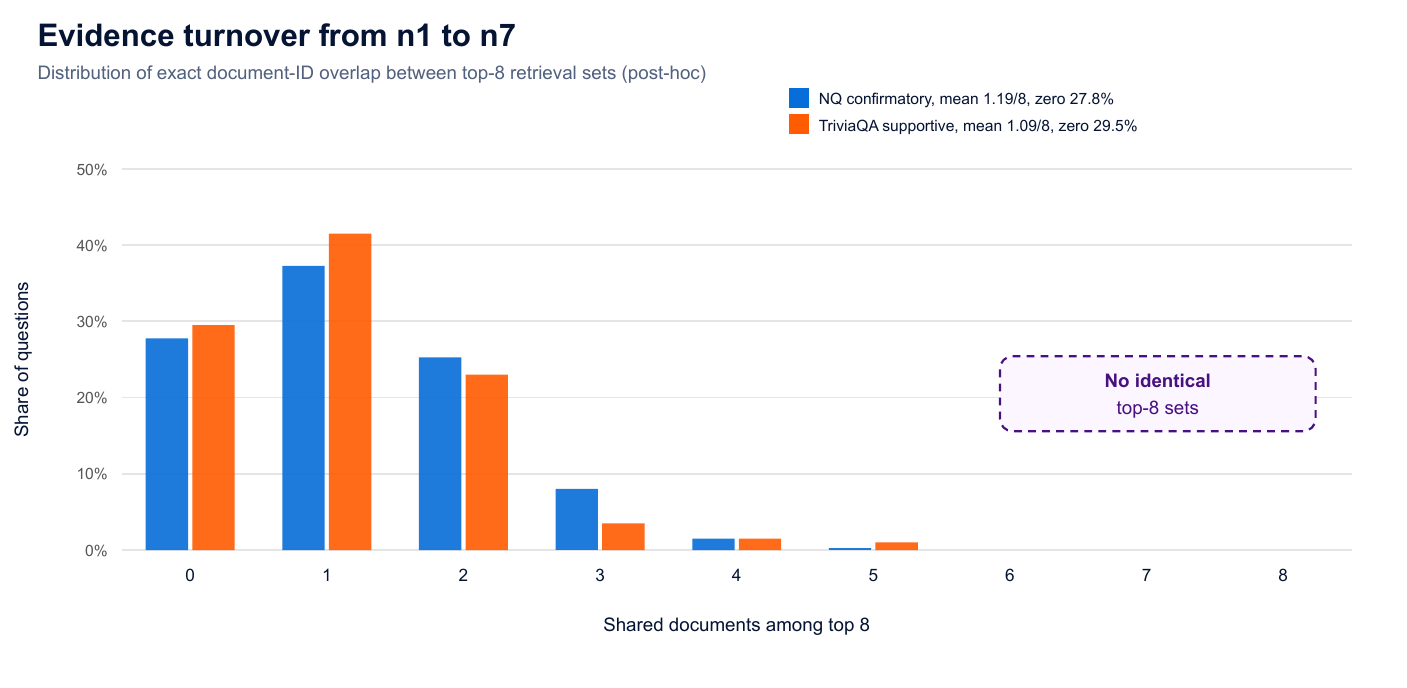}
  \caption{Post-hoc distribution of exact document-ID overlap between the $\none$ and $\nseven$ top-eight retrieval sets.  Large evidence turnover coexists with high cross-scale semantic agreement.}
  \Description{A grouped histogram shows that zero or one shared document is the most common outcome for both benchmarks. Natural Questions shares 1.19 documents on average and TriviaQA 1.09; 27.8 and 29.5 percent of questions, respectively, share no document.}
  \label{fig:overlap}
\end{figure*}

Evidence changes substantially at the endpoint, as shown in Figure~\ref{fig:overlap}. The mean exact document-ID overlap between the two top-eight sets is 1.19/8 for NQ and 1.09/8 for TriviaQA. The corresponding zero-overlap rates are 27.75\% and 29.50\%, and no question has an identical top-eight set. Yet cross-scale semantic agreement remains 78.88\% and 94.63\%. Most answers remain semantically stable despite this turnover, while $\Dhat$ isolates the smaller residual shift beyond repeat noise. Flip questions have lower mean overlap than nonflips (0.825 versus 1.231 documents for NQ), but overlap-bin effects are not monotone and Fisher tests are inconclusive. We treat this only as a mechanism hint.

Absolute agreement and the corrected gap answer different questions. Cross-scale semantic agreement of 78.88\% on NQ and 94.63\% on TriviaQA means that most answers survive severe top-eight turnover. Positive $\Dhat$ says the residual disagreement is nevertheless larger than the same-state baseline. Reporting only agreement would make the snapshots appear interchangeable. Reporting only $\Dhat$ could make a 10.25-pp gap sound as if every tenth question deterministically changed. The strict flip diagnostic locates one interpretable part of the gap but can miss diffuse probability shifts. An audit should report absolute cross-state agreement, baseline-corrected $\Dhat$, and question-level diagnostics.

\subsection{Answer Trajectories Across Scales}
\label{sec:trajectory}

The primary contrast uses only $\none$ and $\nseven$, but the locked 28-pair labels permit a post-hoc view of every adjacent step. Strict stable flips occur at every observed transition. Across the three adjacent NQ steps, the counts are 33/400, 29/400, and 26/400, compared with 40/400 for the endpoint. TriviaQA has 3/200, 5/200, and 2/200 across adjacent steps, compared with 5/200 for the endpoint. The corresponding excess-answer-churn estimates are positive at every step in Table~\ref{tab:ladder}. Churn is therefore not confined to the largest endpoint contrast, although one fixed path cannot establish a general dose-response law or predict arbitrary index updates.

\begin{table}[t]
  \caption{Post-hoc trajectory decomposition. Semantic $\Dhat$ is in pp, and flips use the strict all-four-cross-pairs definition.}
  \label{tab:ladder}
  \small
  \begin{tabular}{llrr}
    \toprule
    \rowcolor{auditheader}
    \textcolor{auditnavy}{\textbf{Study}} & \textcolor{auditnavy}{\textbf{Transition}} & \textcolor{auditnavy}{\textbf{$\Dhat$}} & \textcolor{auditnavy}{\textbf{Stable flips}} \\
    \midrule
    NQ & $\nzero\to\none$ & 10.563 & 33/400 (8.25\%) \\
    \rowcolor{auditstripe}
    NQ & $\none\to\nthree$ & 7.938 & 29/400 (7.25\%) \\
    NQ & $\nthree\to\nseven$ & 6.313 & 26/400 (6.50\%) \\
    \rowcolor{auditstripe}
    NQ & $\none\to\nseven$ & 10.250 & 40/400 (10.00\%) \\
    \midrule
    TriviaQA & $\nzero\to\none$ & 1.500 & 3/200 (1.50\%) \\
    \rowcolor{auditstripe}
    TriviaQA & $\none\to\nthree$ & 2.125 & 5/200 (2.50\%) \\
    TriviaQA & $\nthree\to\nseven$ & 0.875 & 2/200 (1.00\%) \\
    \rowcolor{auditstripe}
    TriviaQA & $\none\to\nseven$ & 2.125 & 5/200 (2.50\%) \\
    \bottomrule
  \end{tabular}
\end{table}

The NQ union over three adjacent flip sets contains 59 questions. Twelve flip at both $\none\to\nthree$ and $\nthree\to\nseven$; for six of them, $\none$ and $\nseven$ return to the same semantic regime, yielding an $X\to Y\to X$ trajectory. Among the 40 endpoint flips, the intermediate $\nthree$ responses strictly align with $\none$ in 8 cases, align with $\nseven$ in 12, form a stable third answer in 6, and disagree across repeats in 12. Two additional questions have mixed or non-transitive pair labels and are reported separately rather than forced into a class. Endpoint-flip questions are enriched for $\nthree$ repeat disagreement (30.0\% versus 8.89\% among other NQ questions). This enrichment is consistent with an answer-fragile subset but does not establish a universal ``unstable middle'' scale.

The no-retrieval output is a descriptive closed-book anchor, not a direct measurement of parametric memory. Of the 40 NQ endpoint flips, 29 have a repeat-stable $\nzero$ answer. Among them, 11 align only with $\none$, 10 only with $\nseven$, and 8 with neither endpoint. There is no dominant one-way movement relative to closed-book behavior. This corpus-scaling intervention complements controlled knowledge-conflict studies but does not identify which evidence or internal memory caused a regime selection.

\subsection{Cross-Family Semantic-Judge Audit}

On the outcome-blind 50-question audit sample, the DeepSeek and OpenAI judges agree on 1,340/1,400 pair labels (95.71\%), with Cohen's $\kappa=0.855$ \cite{cohen1960kappa} and 43/50 exactly matching 28-pair partitions. Question-cluster bootstrap intervals are [91.86, 98.86] for agreement and [0.709, 0.959] for $\kappa$. NQ agreement is 93.70\% ($\kappa=0.827$), and TriviaQA agreement is 100\%. On the same sample, semantic excess churn is 9.50 pp with the original judge and 11.00 pp with the cross-family judge. The robustness audit therefore argues against the effect being produced by same-family judging alone. It does not replace human annotation, and the raw execution trace was not retained. The locked judgment ledgers and recorded zero-tool metadata are hash-bound, and all reported audit statistics were independently recomputed from the ledgers.

\section{Implications and Limitations}
\label{sec:discussion}

\subsection{The Snapshot Compatibility Audit}

From a release-engineering perspective, an index rebuild can be a behavioral release even when model weights and the public API do not change. Old and new retrieval-augmented snapshots should therefore receive the same compatibility scrutiny as two model versions \cite{fard2016churn,traeuble2021backward,caciolai2023regression}. The Snapshot Compatibility Audit organizes this check into four stages.
\begin{enumerate}
\item \textbf{Freeze the comparison.} Select queries without outcome filtering; name the old and new snapshots; and lock the prompt, all exposed generation controls, retrieval depth, rendered evidence, and gold-access boundary. If retrieval is stochastic, either lock its output, as we do, or include its variation in the repeated system state.
\item \textbf{Replicate at question level.}  Collect at least two independent responses per query and state, counterbalancing execution order when shared service conditions may drift. Preserve retries and accepted failures such as \texttt{UNKNOWN}. Selective redraws invalidate the noise baseline.
\item \textbf{Compare blindly.}  Report exact and semantic within/cross agreement, $\Dhat$, and question-cluster intervals. Semantic comparison should hide state, evidence, and correctness. A second judge family or human subset can measure construct sensitivity, but observed outcomes must not determine a replacement.
\item \textbf{Triage, then decide.} Alongside average utility, inspect repeat-stable flips, EM-match/nonmatch transitions, and retrieval overlap. Manually review high-impact changes. The threshold must be application-specific. Positive excess churn establishes incompatibility but does not by itself establish factual harm or justify blocking deployment.
\end{enumerate}

The output-only protocol could audit an index refresh, retriever replacement, chunking or deduplication change, or a change to external memory. For every retrieval-affecting release, we recommend reporting excess answer churn alongside utility. These applications remain untested, and our empirical treatment is limited to one fixed nested FineWeb prefix.

Answer changes are not automatically regressions or harms. The Jill Bigelow answer improves from \texttt{UNKNOWN} to the EM reference, and ambiguous queries can support alternatives, so deployment thresholds require application-specific review.

\subsection{Threats to Validity and Claim Boundary}

\paragraph{Treatment and mechanism.} The identified contrast is the total behavioral difference between two accessible index states on one frozen path, $\none\subset\nthree\subset\nseven$.  Added shards can introduce new top-eight documents and displace documents retrieved at the smaller state; both are part of the treatment.  Because shard order is fixed, nominal scale is confounded with the identities and ranking effects of documents added on this path.  We neither randomize shard subsets nor observe enough independent paths to estimate a content-averaged scaling effect.  Evidence overlap and trajectories expose possible mediators but do not identify which document caused a changed answer.  The results establish that this expansion can alter behavior, not a monotone or universal law assigning a churn rate to ``more corpus.''

\paragraph{Construct validity.} Normalized exact agreement can overcount stylistic changes, while a semantic judge can merge distinct answers or split acceptable aliases. We therefore report both. The larger NQ semantic effect, blind all-pairs design, and cross-family audit make formatting-only and same-family-only explanations unlikely, but the 50-question audit is not human validation. In a narrow post-hoc style sanity check, every normalization-identical endpoint pair in both benchmarks received a semantic-same label within and across snapshots, including every raw-different normalization-equivalent pair, but this does not exclude broader style leakage. Pairwise labels are occasionally non-transitive; we retain rather than repair them after seeing outcomes. Repeat-stable means only within-state semantic agreement across two calls. It is not verbatim stability, calibrated confidence, factual correctness, or user harm. Likewise, EM-nonmatch is not synonymous with falsehood because a verbose answer containing a reference may fail full-string EM.

\paragraph{Sampling and stochasticity.} Each estimate is conditional on its realized serving interface, and same-snapshot repeats estimate that interface's stochastic baseline. The bootstrap treats questions as the independent sampling units and preserves all repeated outputs and pair labels within each unit.  It quantifies question-sampling uncertainty for the frozen cohorts, not uncertainty over alternative corpus partitions, prompts, models, or deployment dates.  Two responses per state are the minimum design that permits a within-state baseline, but give a coarse view of each conditional output distribution; more repeats would reduce contribution-level variance. Only NQ is confirmatory. The separately preregistered 200-question TriviaQA cohort remains supportive and unpooled.

\paragraph{External validity.} We study two DeepSeek generator and serving configurations, one search service, one FineWeb index, top-eight evidence, and two English open-domain QA benchmarks. Both V4-Flash benchmark studies are preregistered, while the V4-Pro subset is post hoc. The V4-Pro replication changes model and transport together, so it does not isolate a model effect or replace cross-family generator evidence. Generic FineWeb retrieval lowers average EM relative to no retrieval in the main setup, so the work neither demonstrates a utility benefit from retrieval nor characterizes an optimized production RAG stack. Replication across model families, retrievers, task types, random corpus paths, and temporal index refreshes is needed before assigning a general prevalence or scaling curve to the phenomenon.

\section{Conclusion}

We asked whether a fixed retrieval-augmented QA system remains behaviorally compatible with its earlier snapshot when only corpus access grows. Accuracy hides offsetting changes, while generation noise confounds one-shot comparisons. The Snapshot Compatibility Audit subtracts cross-snapshot agreement from within-snapshot agreement to estimate excess answer churn beyond ordinary generator variation.

On a fixed DeepSeek QA system whose nested FineWeb prefix grew from one shard to seven, preregistered NQ normalized-exact and semantic excess churn were 6.44 and 10.25 pp, with one-sided 95\% lower bounds of 4.56 and 7.69 pp, while EM changed by $-1.50$ pp. Separately preregistered TriviaQA found supportive excess churn of 3.00 and 2.125 pp while EM moved by $+1.25$ pp. The outcome-blind post-hoc V4-Pro NQ-100 subset found exact and semantic excess churn of 7.25 and 8.75 pp while EM moved by $+3.00$ pp. Stable semantic flips occurred on 40/400 NQ, 5/200 TriviaQA, and 6/100 V4-Pro questions.

Endpoint top-eight sets shared only 1.19 NQ and 1.09 TriviaQA documents on average, yet most answers survived the turnover while a concentrated subset moved beyond repeat noise. On an outcome-blind sample, a cross-family judge agreed with 95.71\% of the original V4-Flash study labels and yielded comparable churn. These small EM shifts did not establish answer-level compatibility.

The evidence comes from one generator family, two serving configurations, one search service, one fixed FineWeb path, and two English QA benchmarks. It does not establish a universal scaling law or document-level cause, and human semantic validation remains open. Even so, positive churn persists across negative, positive, and zero endpoint EM point estimates in the main and matched-subset analyses. Corpus access can therefore alter which answers a system returns without a reliable utility warning. Release tests should report excess answer churn alongside utility.

\section*{Ethical Considerations}

The study uses public QA benchmarks and a web-derived corpus without recruiting participants or inferring user attributes. Natural Questions contains anonymized queries, while web text and model outputs may contain errors, bias, or sensitive material. We release only permitted identifiers and outputs, excluding benchmark text and gold aliases. Excess churn is not a proxy for harm.

\bibliographystyle{ACM-Reference-Format}
\bibliography{references}

\end{document}